\documentclass[sigconf,9pt]{acmart}
\usepackage{caption}
\usepackage{booktabs}

\AtBeginDocument{%
  }

\setcopyright{acmlicensed}
\copyrightyear{2026}
\acmYear{2026}
\acmDOI{XXXXXXX.XXXXXXX}
\acmConference[]{}{}{}

\acmISBN{}

\begin{document}

\title{NL4ST: A \underline{N}atural \underline{L}anguage Query Tool for \underline{S}patio-\underline{T}emporal Databases}






\author{Xieyang Wang, Mengyi Liu, Weijia Yi, Jianqiu Xu$^\ast$}

\affiliation{%
  \institution{Nanjing University of Aeronautics and Astronautics}
  \city{Nanjing}
  \country{China}
}
\email{{xieyang, liumengyi, wjyi_x, jianqiu}@nuaa.edu.cn}

\author{Raymond Chi-Wing Wong}
\affiliation{%
  \institution{The Hong Kong University of Science and Technology}
  \city{Hong Kong}
  \country{China}
}
\email{raywong@cse.ust.hk}

\renewcommand{\authors}{Xieyang Wang, Mengyi Liu, Weijia Yi, Jianqiu Xu, and Raymond Chi-Wing Wong}

\renewcommand{\shortauthors}{Xieyang Wang et al.}

\begin{abstract}
The advancement of mobile computing devices and positioning technologies has led to an explosive growth of spatio-temporal data managed in databases. Representative queries over such data include \textit{range queries}, \textit{nearest neighbor queries}, and \textit{join queries}. However, formulating those queries usually requires domain-specific expertise and familiarity with executable query languages, which would be a challenging task for non-expert users. It leads to a great demand for well-supported \underline{n}atural \underline{l}anguage \underline{q}ueries (NLQs) in spatio-temporal databases. To bridge the gap between non-experts and query plans in databases, we present \textit{NL4ST}, an interactive tool that allows users to query spatio-temporal databases in natural language. NL4ST features a three-layer architecture: (i) \textit{knowledge base and corpus for knowledge preparation}, (ii) \textit{natural language understanding for entity linking}, and (iii) \textit{generating physical plans}.  
Our demonstration will showcase how NL4ST provides effective spatio-temporal physical plans, verified by using four real and synthetic datasets.
We make NL4ST online \footnote{\url{https://nl4st.cpolar.top/nl2secondo/}} and provide the demo video at \url{https://youtu.be/-J1R7R5WoqQ}.
\end{abstract}

\maketitle

{
  \renewcommand{\thefootnote}{}
  \footnotetext{$^\ast$Corresponding Author}
}

\vspace{-0.1cm}
\section{Introduction}
In the past decades, although spatio-temporal databases have been extensively investigated in a wide range of applications~\cite{spatialcrowd}, the assumption that users can efficiently formulate accurate queries poses challenges for non-expert users who are unfamiliar with spatio-temporal knowledge and structured database query languages. To this end, there has been an increasing attention paid to providing a \underline{n}atural \underline{l}anguage \underline{i}nterface for \underline{d}ata\underline{b}ases (NLIDB)~\cite{DBLP:journals/pvldb/Katsogiannis-Meimarakis23}.

\begin{figure}[htbp]
  \centering
  \includegraphics[width=\linewidth]{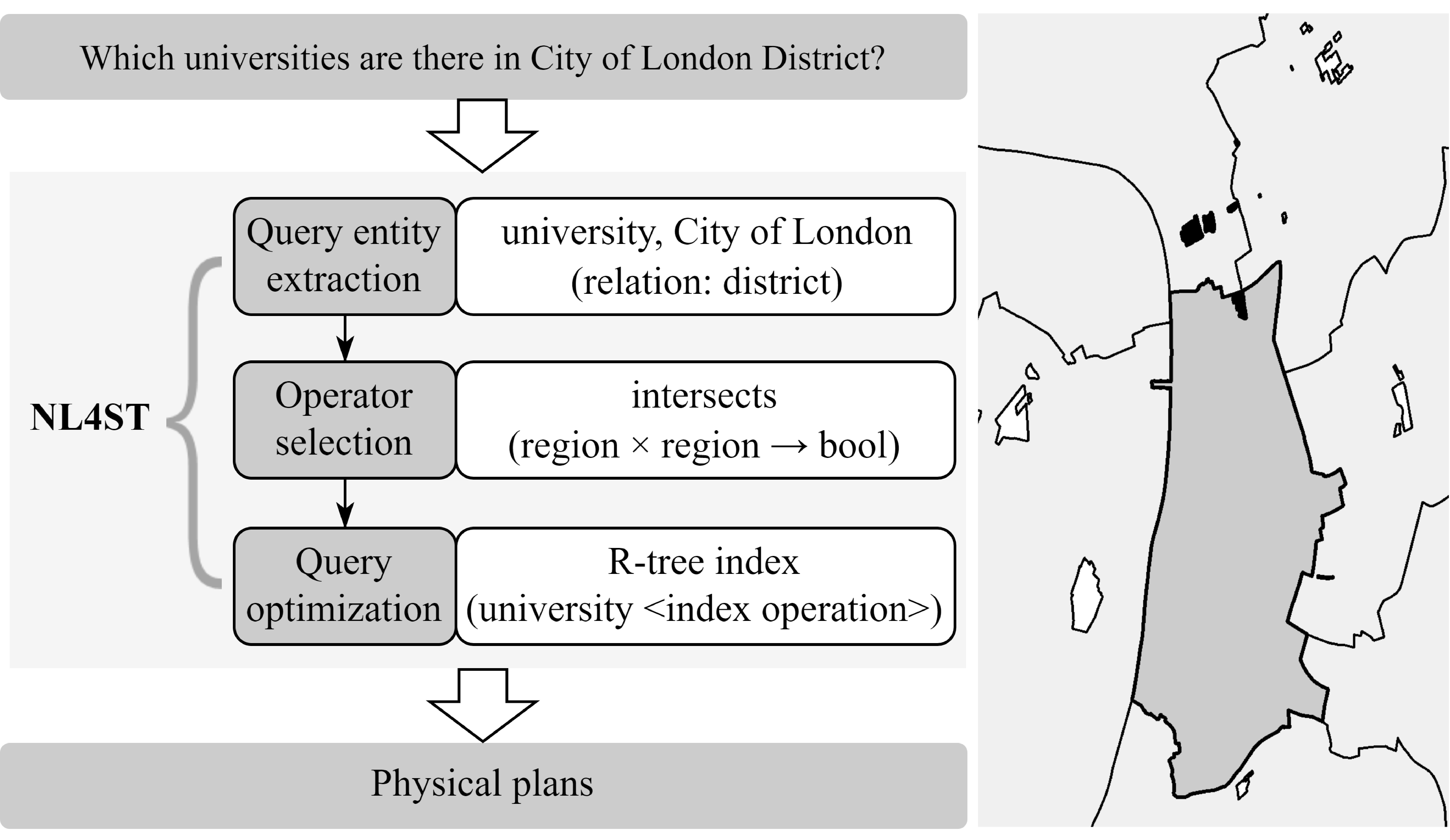}
  \caption{An example of transforming NLQs in spatio-temporal databases.}
  \label{figexample}
  \vspace{-1em}
\end{figure}

In the literature, significant advances have been made for Text-to-SQL~\cite{nalir,GuoZGXLLZ19,opensearch}. However, these approaches primarily focus on relational
databases and treat SQL as the final semantic representation.
From a semantic perspective, SQL lacks native spatio-temporal operators and often leads to ambiguous mappings from \underline{n}atural \underline{l}anguage \underline{q}ueries (NLQs). For example, the phrase ``\textit{in City of London District}'' in Figure \ref{figexample} may map to different spatial predicates in SQL, resulting in ambiguous constructions such as ``\textit{contains (tuple.location)}'' or ``\textit{inside (tuple.location)}''. From an execution perspective, SQL is declarative and hides execution decisions, as operator selection and index usage are deferred to the optimizer. As a result, SQL fails to explicitly capture the execution intent of queries.


In this demo, we investigate \textbf{Text-to-query-plan} for spatio-temporal databases, which directly maps NLQs to query plans, including logical and physical plans. Logical plans describe domain-specific query semantics through algebraic operators, while physical plans specify domain-specific execution operators and index usage. We use the physical plan as the output, which allows the Text-to-query-plan translation to directly capture the semantics of NLQs and explicitly control the execution semantics and efficiency.
Although query plans offer a nuanced approach to data manipulation, Text-to-query-plan presents the following three challenges.

\noindent \textbf{Challenge 1. The Difficulty of Applying Text-to-SQL Methods to Text-to-query-plan.} Existing Text-to-SQL methods rely on \underline{l}arge \underline{l}anguage \underline{m}odels (LLMs), while spatio-temporal query plans require rich domain knowledge and complex executable structures, making training data construction costly. Moreover, current efficiency-control techniques generally assume a black-box context, raising challenges in (i) \textit{domain knowledge alignment} and (ii) \textit{correct and efficient plan generation}.

\noindent \textbf{Challenge 2. Lack of Domain-specific Knowledge.} 
The lack of domain-specific knowledge in LLM-based systems causes hallucinations and incorrect entity alignment. As shown in Figure \ref{figexample}, ``\textit{City of London District}'' is often misaligned by GPT-4o at the column level, although the entity corresponds to a specific value-level representation, highlighting the necessity of domain-aware entity grounding to enhance natural language understanding.

\noindent \textbf{Challenge 3. The Complexity of Query Plan Generation.} 
The large combination space of operators and indexes makes physical plan generation difficult to scale while maintaining efficiency. For the NLQ in Figure \ref{figexample}, GPT-4o employs the \textit{ininterior} operator instead of \textit{intersect} and thus fails to utilize available indexes.

There are limited studies about NLIDB in the spatio-temporal domain ~\cite{nalspatial,nalmo}, as integrating spatial and temporal dimensions in a unified framework introduces complexity.

We develop NL4ST, a \underline{n}atural \underline{l}anguage query tool for \underline{s}patio-\underline{t}emporal databases, that enables users to query spatial, temporal, and trajectory data using natural language. 
In comparison with Text-to-SQL, the approach of Text-to-query-plan (without involving the intermediate result SQL) is able to (i) \textit{avoid ambiguity in text languages} and (ii) \textit{empower the expression by involving operators and data types from specific domains}.
The tool consists of three layers: (i) \textit{knowledge base and corpus}, (ii) \textit{natural language understanding}, and (iii) \textit{generating physical plans}. 
NL4ST generates knowledge bases through database analysis and constructs a corpus by collecting NLQs from published papers on top conferences and journals, which are further augmented by LLMs. The knowledge bases and corpus are leveraged to enable knowledge retrieval and example reference, guiding users in submitting robust queries, which can also be independently utilized for subsequent natural language understanding. 
NL4ST utilizes \underline{n}atural \underline{l}anguage \underline{p}rocessing (NLP) tools for coarse-grained entity extraction, followed by fine-grained entity extraction using an advanced algorithm and the knowledge base. 
NL4ST leverages a query type classification model and structured language models corresponding to query types, which prunes the candidate set for physical plan generation. In addition, NL4ST employs a model to predict the optimal physical plan. The system is published online and can be accessed at \url{https://nl4st.cpolar.top/nl2secondo/}.

\section{System Architecture}
\begin{figure}
  \centering
  \includegraphics[width=0.9\linewidth]{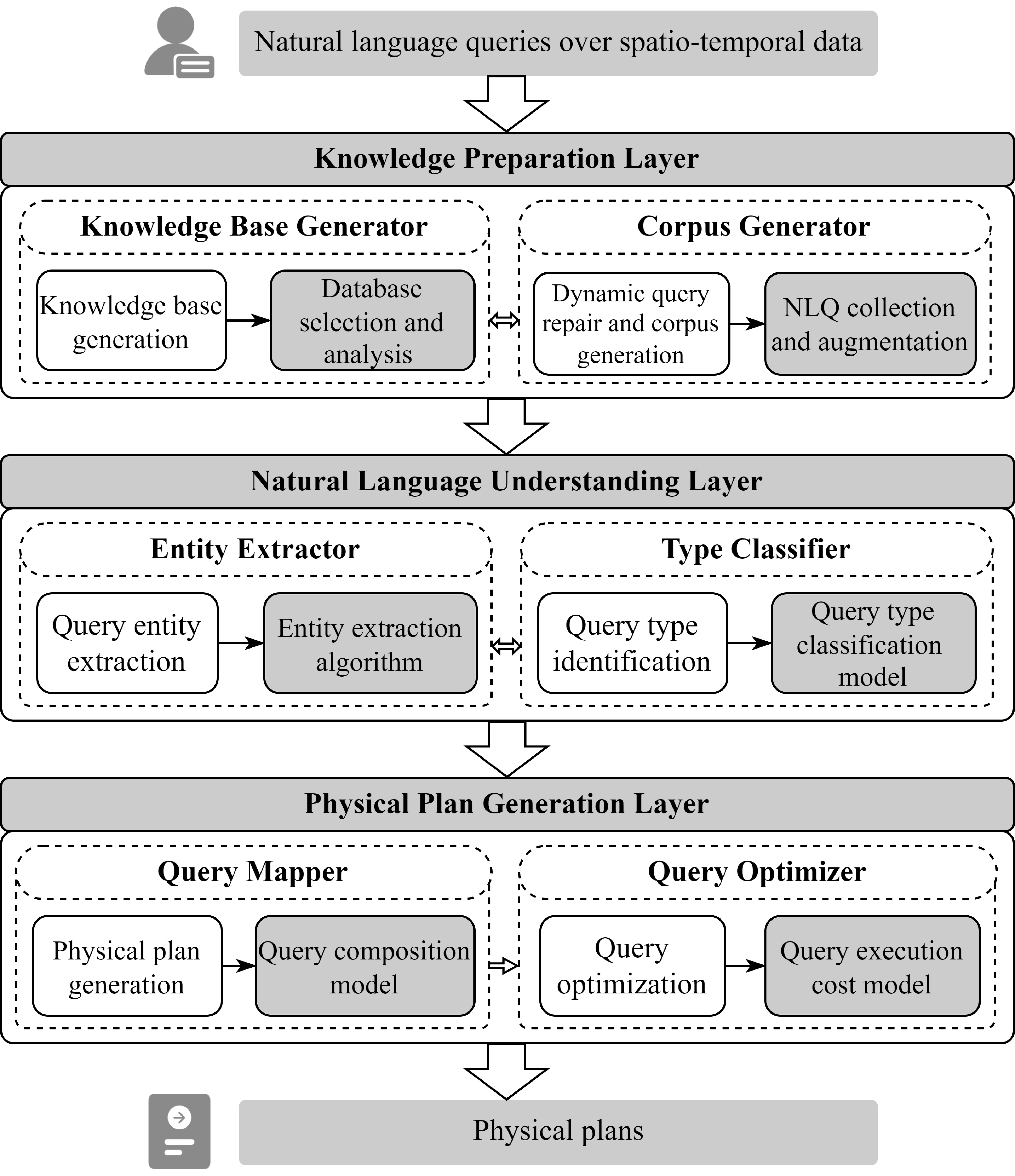}
  \caption{The architecture overview of NL4ST.}
  \label{overview}
  \vspace{-2.0em}
\end{figure}

We provide the framework of NL4ST in Figure \ref{overview}, which adopts a three-layer architecture: (i) \textit{knowledge preparation}, (ii) \textit{natural language understanding}, and (iii) \textit{physical plan generation}.

\subsection{Knowledge Preparation}
NL4ST addresses the lack of spatio-temporal domain knowledge through (i) \textit{the knowledge base generator} and (ii) \textit{the corpus generator}.

\noindent \textbf{Knowledge Base Generator.} 
We construct a spatio-temporal knowledge base to support knowledge retrieval and entity candidate pruning. The knowledge base consists of a relation knowledge base and a location knowledge base. The relation knowledge base stores relational tables with identifiers, names, and spatio-temporal attributes, while the location knowledge base contains objects with \textit{point}, \textit{line}, and \textit{region} representations. The two knowledge bases are tightly coupled, where location entries reference identifiers defined in the relation knowledge base.

\noindent \textbf{Corpus Generator.} 
We construct an NLQ corpus with about 5,000 templates covering \textit{spatial queries} and \textit{moving objects queries} to support user guidance and subsequent query type identification. NLQs are gathered from published papers and augmented by GPT-4o with designed prompts. To ensure alignment with databases, an automatic detection and repair method is employed to refine the NLQs. We also construct templates to enable the generation of a large volume of high-quality NLQs tailored to different databases. The corpus consists of query types and NLQs. The relations and entities in the NLQs are derived from selected databases. 

\subsection{Natural Language Understanding}
NL4ST primarily performs NLP, including two modules: (i) \textit{the entity extractor} and (ii) \textit{the type classifier}.

\noindent \textbf{Entity Extractor.} The extractor aligns the entities in NLQs with database values. \textit{spaCy} is used for initial entity recognition, which is an NLP tool, followed by candidate entity generation into two lists: (i) \textit{the number list}, containing entities labeled with time, number, cardinal, or quantity, and (ii) \textit{the information list}, potentially containing relations, locations, or objects.  
Then, we prune candidates by a fine-grained extraction algorithm to extract the exact key entities. The object identifier and the nearest neighbor identifier encompass the contextual information pertaining to the object and the keyword
like ``\textit{nearest}''.
Leveraging the number and the associated distance unit in the phrase, the precise value of the distance threshold is determined. Besides, we utilize the constructed spatio-temporal knowledge base to perform direct retrieval of relations, objects, and location information in the information list.


\noindent \textbf{Type Classifier.} Given the diversity of spatio-temporal queries, providing a general translation model for all query types is impractical. Identifying and categorizing query types significantly narrows the search space for constructing physical plans. LSTM offers good performance and reduced training time, and thus, we employ LSTM for model training by the given corpus. The trained model can be utilized for type identification.  

\subsection{Physical Plan Generation}
NL4ST advances to the physical plan generation, including two key modules: (i) \textit{the query mapper} and (ii) \textit{the query optimizer}. 

\noindent \textbf{Query Mapper.} We construct the candidate physical plans by combining entities and operators required for different query types through mapping rules, based on a pre-defined query model. We take a moving object nearest neighbor query as an example, limited by space. The operator \textit{knearest} generates a tuple stream that mirrors the structure of the input stream.
Limited by the time when objects arrive among the \textit{k} nearest moving objects, we create a version of \textit{moving trains}, where units are sorted by the start time. We can obtain the physical plan by using the relation \textit{UTOrdered}. The template of the query expression is as follows, and the content in ``⟨⟩'' will be extracted from NLQs. 

\vspace{0.1em}

\textit{query UTOrdered feed filter [(deftime(.UTrip) \textbf{intersects} ⟨ period ⟩)]
\textbf{knearest}[UTrip, ⟨ object ⟩, ⟨ k ⟩]consume;}

\vspace{0.1em}

\noindent \textbf{Query Optimizer.} To further enhance the user-friendliness, we optimize query efficiency by focusing on index utilization. We count and collect the number of records involved in each spatio-temporal relation. Then, we estimate the filter rate based on the operator constraints and the relations involved in the query, where the filter rate refers to the percentage of records that meet specific query criteria relative to the total number of records in the dataset. For queries with a high filter rate, we enumerate candidate physical plans with indexes for queries generated by the query mapper. Subsequently, we employ a cost model to execute the physical plans with sampled small-scale data to predict the temporal overhead of each candidate physical plan, ultimately providing users with the optimal physical plan.


\begin{figure*}[htbp]
  \centering
  \includegraphics[width=0.9\linewidth]{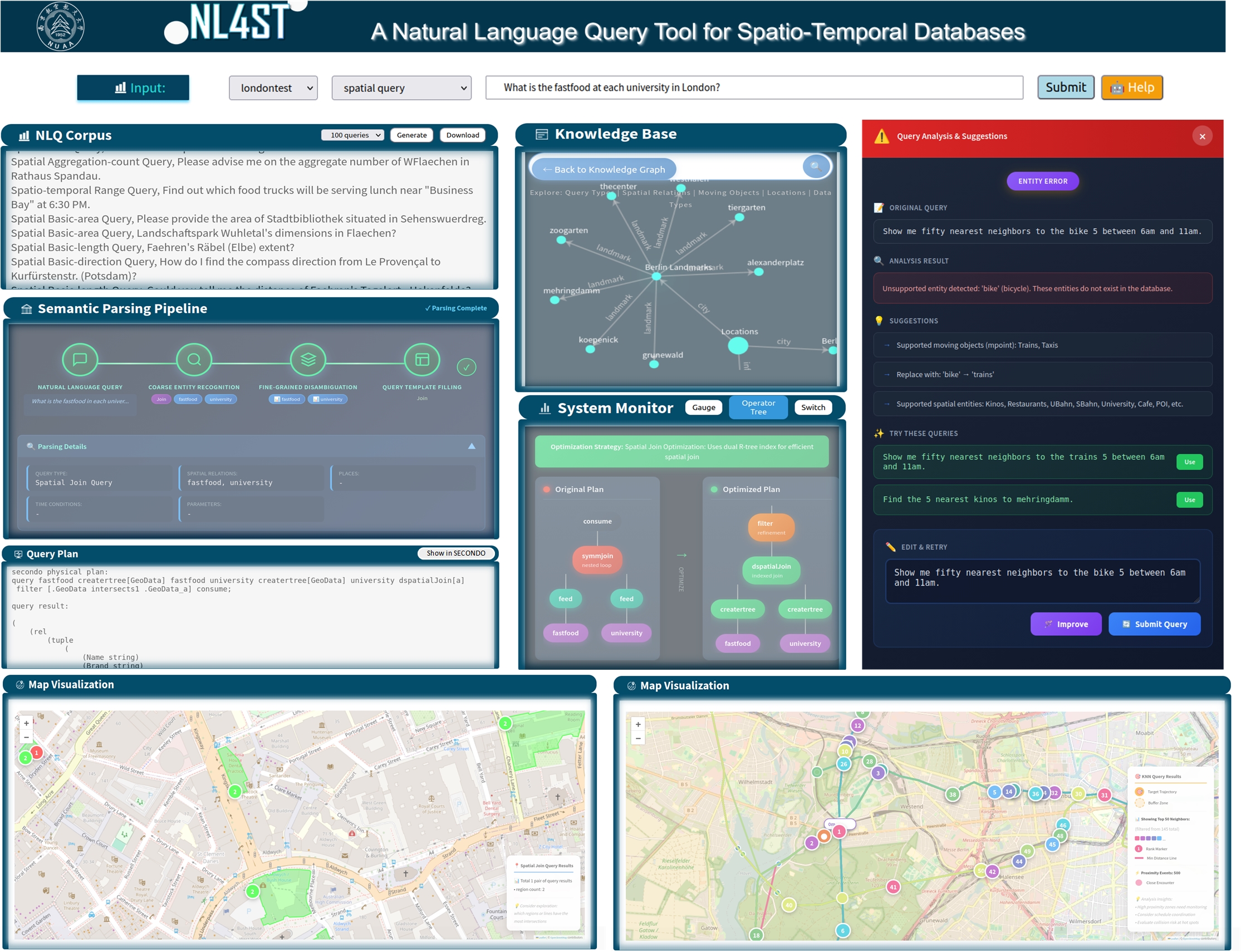}
  \caption{The screenshot of NL4ST.}
  \label{screenshot}
  \vspace{-1.0em}
\end{figure*}

\section{Demonstration}

We deploy NL4ST in a desktop PC (Intel(R) Core(TM) i9-11950H CPU, 2.60 GHz, 32 GB memory, 512 GB disk) running Ubuntu 22.04 (64 bits, kernel version 6.8.0-49-generic). We provide an interactive web UI for users, and the backend is integrated into an extensible database system SECONDO \cite{GutingBD10} as an algebra module. Users perform queries based on the loaded datasets and obtain physical plans as well as visualization results. We leverage four datasets, three of which pertain to urban data\footnote{\url{https://github.com/zhongmove/NL4ST}\label{footnote1}} while the fourth is focused on hydrological data\footnote{\url{https://www.hydrosheds.org/products}\label{footnote2}}. (i) The dataset \textit{nanjingtest} contains taxis, districts, part of roads and
POIs in Nanjing. (ii) The dataset \textit{londontest} includes a subset of POIs and districts in London.
(iii) The dataset \textit{berlintest} encapsulates urban data of Berlin, encompassing trains,
POIs, and rivers. (iv) The dataset \textit{chinawater} stores partial water
systems in China. Detailed statistics are reported in Table \ref{tabdata}. We demonstrate the capacity of NL4ST to transform spatio-temporal NLQs into physical plans, aiming to provide an interactive tool for users
to learn the working mechanisms of the Text-to-query-plan techniques introduced above.

\begin{table}[htbp]
\vspace{-1.0em}
  \centering
  \setlength{\abovecaptionskip}{0cm}
  \caption{Data statistics.}
  \label{tabdata}
  \resizebox{\linewidth}{!}{
  \begin{tabular}{cccccc}
    \toprule
    Datasets & \#tables & \#points  & \#lines & \#regions& \#moving objects\\
    \midrule
    nanjingtest & 6 & 9,000&887 & 13& 147\\
    londontest & 6& 9,032 &9,728& 12,669& 0\\
    berlintest & 50& 3,040 &4,078& 330& 562\\
    chinawater & 2& 0 &8,399& 2,907& 0\\
  \bottomrule
\end{tabular}}
\vspace{-1.0em}
\end{table}

\subsection{Demonstration Scenarios}
Figure \ref{screenshot} presents the screenshot of NL4ST, which enables users to perform (i) \textit{basic spatial query}, (ii) \textit{time interval query}, (iii) \textit{range query}, (iv) \textit{nearest neighbor query}, (v) \textit{join query}, (vi) \textit{similarity query}, and (vii) \textit{aggregation query}. We perform the demonstration in the following steps and provide two query scenarios.

\noindent \textbf{Corpus Generation and Example Reference.} Initially, users can select varying quantities of corpus data for generation and utilize the generated spatio-temporal query corpus to train their own models. These NLQ examples in the corpus can also guide users in formulating high-quality queries.

\noindent \textbf{Knowledge Retrieval.} Subsequently, users can retrieve entities from the pre-constructed knowledge base, which can be utilized for formulating subsequent NLQs with referencing examples.

\noindent \textbf{Database Selection and NLQ Input.} Given that the schema affects the execution of the intended physical plan, users must first select the target database for the query. To simplify this step, users can check sample queries that can be selected and used as input directly. After inputting the NLQ, users can click the ``Submit'' button.


\noindent \textbf{Execution Result Display.} After submission, the tool presents the parsing process. This process first performs coarse-grained entity recognition, followed by fine-grained entity disambiguation, and then conducts detailed entity extraction for query template instantiation. Based on the extracted information, the system generates a physical plan and returns the corresponding query results.

\noindent \textbf{Map Visualization and Interaction.} Users can click the ``Show in SECONDO'' button to view visualization results inside the database. The tool also visualizes the results on an interactive map. For each query type, users can get a tailored visualization design, which also guides users to detect the real-world use cases of the results.

\noindent \textbf{Error Handling.} When the input contains an unsupported query type, a syntax error, or an entity related error, users can click the “Help” button. The tool explains the cause of the error and provides possible corrections. Users can revise the input based on the suggestions and resubmit the query.

\noindent \textbf{Query Efficiency Display.} Finally, the tool reports the execution time of the physical plan. Users can click the ``Switch'' button to compare the execution time of the original query with that of the optimized query. Users can also inspect the physical plan through a visual operator tree
via the ``Operator Tree'' button.

\begin{table}[htbp]
\vspace{-1.0em}
\setlength{\abovecaptionskip}{0cm}
  \caption{Example queries.}

  \label{tabsample}
  \resizebox{\linewidth}{!}{
  \begin{tabular}{ll}
    \toprule
    Example NLQ & Type \\
    \midrule
    Q1: What is the fastfood at each university in London? & Spatial join query\\
    Q2: Show me fifty nearest neighbors to the train 5 between 6am and 11am. & Moving objects nearest neighbor query\\
  \bottomrule
  \vspace{-1.0em}
\end{tabular}}
\vspace{-1.0em}
\end{table}

\noindent \textbf{Scenario 1: Querying Spatial Data.} We consider the spatial join query \textit{Q1} provided in Table \ref{tabsample} and demonstrate the system workflow following the steps described above. Users can first inspect the corpus and
the knowledge base, which provide a set of standard queries and
entities for reference. Users then select a target database and submit an NLQ, optionally choosing from the recommended queries. The detailed semantic parsing process is presented in Figure \ref{screenshot}, based on which the system generates a physical plan, and the query results are visualized on the map.
NL4ST applies an R-tree index for optimization, and users can examine the system monitor to obtain the query execution time. The query execution process is also exposed through a visual operator tree, which can be accessed by clicking the ``Operator Tree'' button.


\noindent \textbf{Scenario 2: Querying Moving Objects.} The nearest neighbor query \textit{Q2} showcased in Table \ref{tabsample} also follows the above steps to generate the results and provide a visualization.
In the \textit{Map Visualization} module, 
NL4ST uses connecting lines to indicate the nearest neighbor order and distance. This visualization guides users to detect real-world use cases, such as identifying locations or time periods where moving objects are in close proximity.
If existing issues, such as entity related errors as shown in Figure \ref{screenshot}, users can click the ``Help'' button to obtain error explanations and guidance. In the analysis result area, users can inspect detailed error messages, while the suggestions area provides potential corrections. Users can also directly try the recommended queries listed.


\subsection{Evaluation Metrics}
We evaluate NL4ST using three metrics: (i) \textit{response time}, (ii) \textit{translatability}, and (iii) \textit{translation precision}. Translatability measures the fraction of NLQs that can be translated into physical plans, while translation precision measures the fraction of translated physical plans that produce expected results. The average response time, translatability, and translation precision of NL4ST are 1.9 seconds, 93\%, and 90\%, respectively.

\bibliographystyle{ACM-Reference-Format}
\bibliography{sample-base}


\end{document}